# Neural network inverse design of nanophotonic scintillators

**Nathan Regev[1], Avner Shultzman[1], Francis Loignon-Houle[2], Charles Roques-Carmes[3, 4], and Ido Kaminer[1]**

[1]*Technion – Israel Institute of Technology, Haifa 32000, Israel*

[2] *Institute for Instrumentation in Molecular Imaging (I3M), CSIC-Universitat Politècnica de València, Valencia 46022, Spain*

[3]*E.L. Ginzton Laboratory, Stanford University, Stanford 94305 CA, USA*

[4]*Institute of Science and Technology Austria (ISTA), Am Campus 1, Klosterneuburg 3400, Austria*

**Abstract:**

Scintillators are materials converting high-energy radiation into optical light, essential in a range of technologies such as medical imaging systems and security scanners. Scintillator development and optimization have remained limited by the complexity of their underlying physics, involving stochastic cascades of electron-electron, electron-phonon, and electron-photon interactions. Such processes are typically modeled by non-differentiable Monte Carlo simulations, limiting the applicability of machine learning for scintillator development. Here we present a physics-informed neural network that learns the scintillation cascade process from the incident high-energy particle to photon emission, substantially accelerating scintillator design and optimization. Combining this neural network with photonic simulations enables end-to-end differentiable optimization of the scintillator geometry. This allows us to optimize for arbitrary figures of merit, such as specific target emission patterns.. We demonstrate the concept and characterize it relative to previous approaches by inverse design of nanophotonic scintillators for X-ray imaging.

Scintillation, the emission of light by materials exposed to high-energy particles, is vital for applications across multiple areas of science and technology. When coupled with a photodetector, scintillators serve as sensors for high-energy particles. Scintillators are fundamental to modern medical diagnostics[1,2], such as positron emission tomography (PET) and computed tomography (CT). Furthermore, scintillators are essential in frontier research applications, including high-energy physics, space exploration and crystallography. Each particle excites energetic electrons that undergo a cascade of interactions forming electron-hole pairs, which recombine to emit light through spontaneous emission. Due to the complex physics of scintillators, their improvement requires balancing multiple material properties that are in inherent conflict: high stopping power, efficient light emission, and optical transparency[3]. High stopping power demands dense, high atomic number materials to efficiently absorb incoming particles, but such materials often introduce non-radiative losses and lack optical transparency. Optical transparency is typically associated with lower-density materials that reduce particle absorption, essential for emitted light to escape. Meanwhile, efficient light emission requires clean electronic structures and minimal defects, which can be compromised in the dense scintillating materials required for high stopping power.

To surpass these inherent conflicts, the next frontier in radiation detection aims for meta-scintillators[4] that utilize nanoscale engineering of structural and material properties. One prominent approach is nanophotonic scintillators that directly enhances the light emission. This is done by manipulating the material's geometry at the wavelength scale and below [5]. Two main design strategies have emerged for integrating nanophotonics with scintillators[6]. Works by Lecoq[7] applied a photonic crystal coating to the scintillator, thereby enhancing the extraction of the scintillation light from the material's interior. Using such subwavelength periodic nanoscale structures, an order of magnitude in light amplification was later achieved by Roques-Carmes et al[8]. Another approach, introduced by Kurman et al.[9], shapes the interior scintillating material to amplify the photonic density of states, enhancing scintillation by accelerating the intrinsic spontaneous emission process, as recently demonstrated with photonic crystals[10] and plasmonic materials[11]. Such design strategies allow photonic enhancements by the Purcell effect[12] as well as improving emission lifetime[13,14].

In parallel, hybrid scintillator architectures have been proposed to address the trade-off between efficient high-energy particle absorption and light emission. This approach leverages the combination of disparate materials, where a high-Z component provides the necessary stopping power and a second material performs fast electron-hole recombination[15–18]. Further advancements were made by incorporating nano-materials like quantum dots[19–25] as a scintillating material, to engineer a desired scintillation response like high scintillation efficiency, low reabsorption rates and spectral tunability.

Developing next-generation meta-scintillators is currently hindered by the absence of efficient differentiable modeling and design tools. The multiscale, stochastic nature of scintillation—encompassing initial interaction (e.g., photoelectric absorption, Compton scattering), electron cascades, and visible photon emission[3,26]—renders quantitative simulation computationally intensive. While Monte Carlo (MC) methods are the standard for sampling these complex realizations, their non-differentiable nature limits their utility in gradient-based inverse design. Recent advances have sought to bridge this gap: Roques-Carmes et al[8]. integrated MC simulations with optical propagation tools to model nanophotonic scintillators, while Shultzman et al[27]. introduced an inverse design framework using analytical approximations of radiation absorption.

Simultaneously, emerging efforts such as OptiGAN[28] have focused on training surrogate neural networks to accelerate Monte Carlo simulations for bulk scintillators. Existing methods provide MC simulation acceleration but cannot support active optimization. To guide future scintillators design and optimization requires capturing the underlying complex, nanoscale interplay between desired structural geometries and intrinsic radiation transport dynamics—the critical gap that our approach directly addresses.

Here, we present scintillator physics-informed neural network (SPINN), a machine learning methodology for the end-to-end optimization of next-generation scintillators. The framework combines a differentiable model of the scintillation process with an analytical light-propagation simulator, enabling gradient-based optimization of scintillator geometries and optical response that follows the X-ray-to-optical conversion process. Central to this approach is a neural network module that learns the MC simulation of radiation–matter interactions while remaining fully differentiable.

To realize this capability, we employ a physics-informed neural network (PINN) surrogate that learns the mapping between the scintillator geometry and the resulting optical emission. By replacing the stochastic MC simulator, the PINN captures the underlying structural dependencies of the conversion process. Crucially, it enables the calculation of emission gradients with respect to the geometry—a capability missing from existing solutions, yet essential for executing an inverse design approach on such structures. We demonstrate this PINN-enhanced inverse-design methodology on the optimization of nanophotonic scintillators of layered geometry. More generally, the same framework can be extended to other geometries and other high-energy particles, including γ-rays, protons, and neutrons, through the incorporation of the corresponding interaction models.

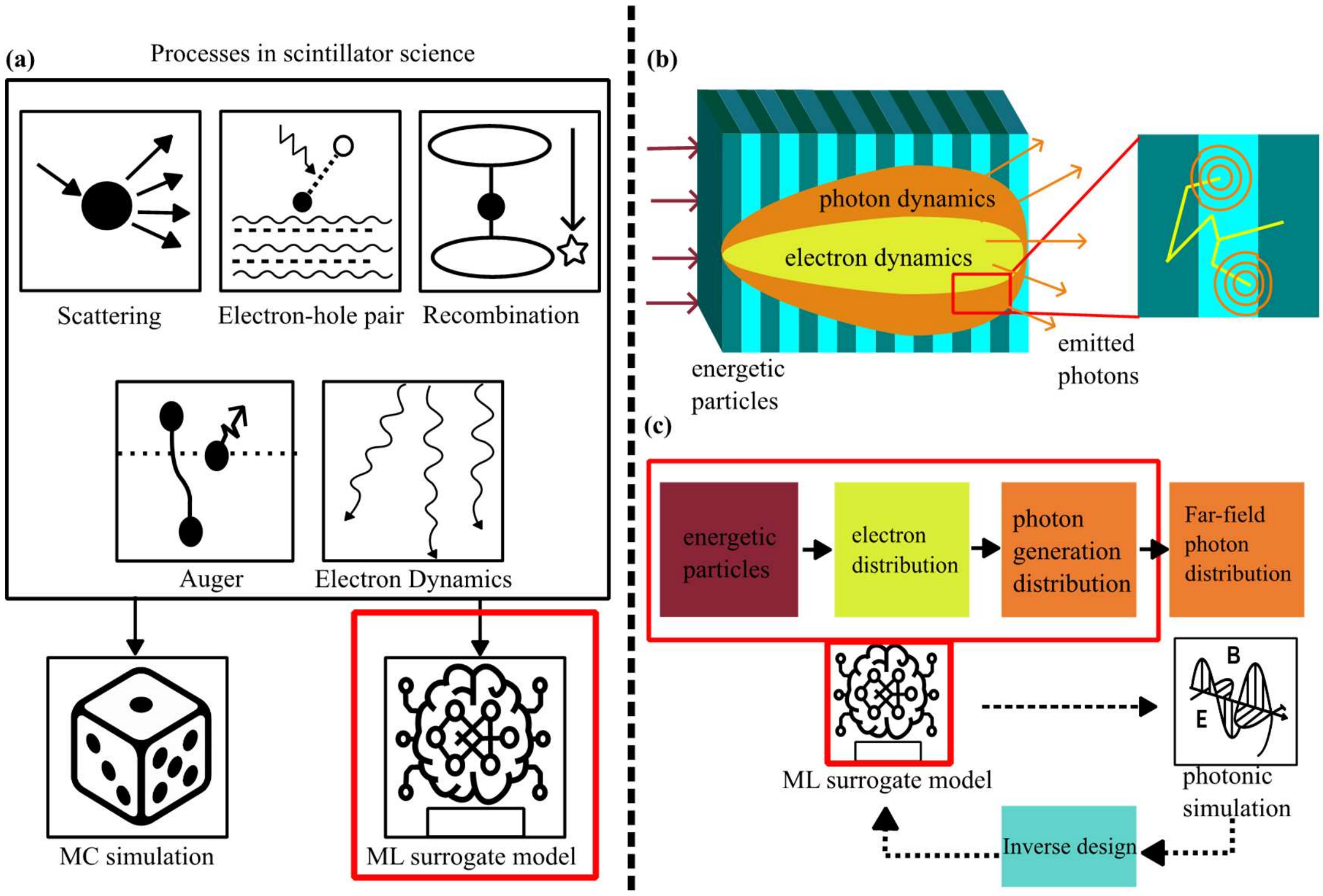

**Figure 1: Physics-informed neural network (PINN) enabling inverse design of scintillators. (a)** A wide range of physical processes are triggered following the interaction of an X-ray photon or other energetic particles with matter – all part of the complexity of scintillator science. These processes are typically modeled by probabilistic Monte Carlo (MC) simulations such as Geant4[29]. The complexity of these processes and their probabilistic simulations present a major obstacle for machine learning–based modeling of scintillators. We propose a PINN for the scintillation process. **(b)** Visualization of the underlying scintillation process: First, electron dynamics: an incident energetic particle scatter and excites energetic electrons that travel through the material, depositing energy via a cascade of scattering events. Part of this energy can lead to the emission of optical photons if reaching a luminescence center (zoom-in panel). Second, photon dynamics: the emitted optical photons propagate according to Maxwell's equations, coupling out of the scintillator and providing the desired signal. We demonstrate the concept of PINN surrogate inverse design on a nanophotonic scintillator made from a layered nanostructure with alternating refractive indices. Successful optimization leads to directional emission (orange arrows). **(c)** SPINN architecture: Integrating a PINN surrogate model creates a differentiable link between the scintillator's layer structure and the underlying electron dynamics. Combining this surrogate model with a photonic simulator, which is already differentiable, enables us to incorporate machine learning tools. This enables the inverse design process (marked in teal); the full optimization process is detailed in Figure 2.

PINNs have gained prominence in recent years[30–33]. Such neural networks are trained on data from a physical process by imposing the governing equations and boundary/initial conditions as loss terms or as explicit prior to the model. Once trained, the model serves as a fast, physics-consistent surrogate for simulation and design. Compared with direct optimization on forward solvers, surrogate-based optimization offers substantially simpler and faster evaluation during large design sweeps at the cost of up-front training. Importantly for many optimization algorithms, including ours here, the PINN model provides end-to-end differentiability, enabling gradient optimization even when the original forward model includes non-differentiable steps such as remeshing, limiters, or MC noise. These advantages have motivated applications of other PINNs in areas such as aerodynamic shape design[34], proton beam-induced energy transfer[35], and radiative heat-transfer modeling[36].

Introducing PINN surrogates to scintillators science allows tailoring the structure to achieve improved far-field scintillation emission profiles, addressing the longstanding trade-off between resolution and efficiency in scintillator design (Figure 1). While our focus is on nanophotonic geometries, the core methodology can be extended to the design and optimization of other types of meta-scintillators, opening a path toward machine-learning driven development of next-generation radiation detectors.

**Scintillator Physics-Informed Neural Network (SPINN)**

Finding the optimal structure poses a challenging problem, due to the stochastic dependence of the output light signal on the scintillator geometry. To demonstrate the approach, we consider multilayer stack with subwavelength scale thickness geometries, which can be fabricated[37] and have been shown to significantly enhance the light signal[16]. Previous works[27] maximized the photonic emission dynamics but did not account for the internal electron dynamics governing scintillation efficiency and rate. While such model is sufficient for bulk scintillators where thickness exceeds the electron mean-free path, it fails for emerging materials like quantum dot or nanophotonic scintillators. In these systems, hot electrons traverse material interfaces—often moving counter to the incident radiation—rendering macroscopic linear yield models inaccurate and necessitating a high-fidelity transport model. Here, the proposed surrogate architecture captures this relation between the signal and the structure's geometry, and integrates it into the inverse design, while accounting for all physical processes.

The detected light yield signal produced by the nanolayered structure can be described by the following expression, integrating the light from all emitters' positions and orientations:

$$S(\vec{d},\theta,\omega) = \int \Gamma_{\text{eff}}(\theta,\omega,z,\vec{d})Y(\omega)G(z,\vec{d})dz, \tag{1}$$

where the position along the axis of the photonic crystal is z, the vector of layers thicknesses is $\vec{d}$, the emission frequency is $\omega$, the angle of emission with respect the axis of the photonic crystal is $\theta$, the Purcell enhancement factor is $\Gamma_{\text{eff}}(\theta,\omega,z,\vec{d})$, the spectrum of the emission is $Y(\omega)$, and the spatial distribution of generated photons $G(z,\vec{d})$. Both $Y(\omega)$ and $\Gamma_{\text{eff}}(\theta,\omega,z,\vec{d})$ are deterministic variable, while $G(z,\vec{d})$ is s a random variable estimated via MC simulations or surrogate models. The optimization process utilizes the expected value of this random variable, approximated by averaging multiple realizations from the estimation tool.
For optimizing a scintillator, different figures of merit can be employed. For example, to maximize the signal's yield while maintaining the sample thickness at a constant thickness $c = |\vec{d}| = \sum_i d_i$, we can choose the objective function to be $L(\vec{d}) = -\int_{\theta\in[\bar{\theta}]} \int S(\vec{d},\theta,\omega)\, d\omega d\theta$, integrated over the spectrum and over the acceptance cone of the detector $\bar{\theta}$. Then,

$$\vec{d}_{\text{optimal}} = \underset{d}{\operatorname{argmin}}\, L(\vec{d}) \text{ subject to } |\vec{d}| = c. \tag{2}$$

Other objective functions and constraints can be considered, such as ones optimizing angular distribution $\Theta(\theta)$ and spectrum $\Omega(\omega)$ for the emission. In this case, the objective function is directly expanded to $L(\vec{d}) = -\int_{\theta\in[\bar{\theta}]} \int S(\vec{d}, \theta, \omega)\Theta(\theta)\Omega(\omega)\, d\omega d\theta$.

Previous optimization schemes[27,37] approximated the spatial distribution of generated photons $G(z, \vec{d})$ using an exponential decay model[38], following the absorption profile of energetic particles. This assumption holds for bulk scintillators since their thickness is much larger than the typical electron travel distance; however, it breaks down for many meta scintillators with nanoscale features, thus breaking the exponential decay model of $G(z, \vec{d})$ and necessitating a better model.

In the cases analyzed here, layer thicknesses are on the order of the optical wavelength and comparable to the mean-free path of the excited hot electrons. While the nanophotonic response and specifically the far-field amplification is a function of the emission depth within the layered geometry, the mean free path introduces a critical transport effect. Because this distance is commensurate with the layer thickness, electrons generated by X-ray absorption in one layer can readily traverse into adjacent layers before undergoing radiative recombination. Consequently, the final light emission profile is not strictly governed by the local X-ray absorption of a single material but is instead shaped by the electron contributions of the surrounding layers. This non-local generation process breaks the conventional exponential approximation and necessitates a more integrated approach when optimizing the structure for maximum light yield.

Accurate modeling of $G(z, \vec{d})$ and the emitters distribution in many scintillators relies on capturing the complex underlying electron dynamics triggered by the incident radiation[39]. Although each step in the particles' transport can be analytically modeled with corresponding interaction cross sections, they cannot be combined in a closed analytical formulation due to the stochastic nature of the interactions. As a result, the modeling is done using stochastic MC[29,39,40] simulations like the Geant4 toolkit that we apply here to simulate the X-ray absorption and transport within nanophotonic scintillators, accounting for particle transport, scattering and energy deposition. However, the non-differentiable nature and high computational cost of MC simulations make them incompatible with the gradient-based nature of inverse design approaches. As a result, current scintillator optimization often ignores these critical physical processes, limiting their ability to fully optimize the scintillation process from particle absorption to photon detections.

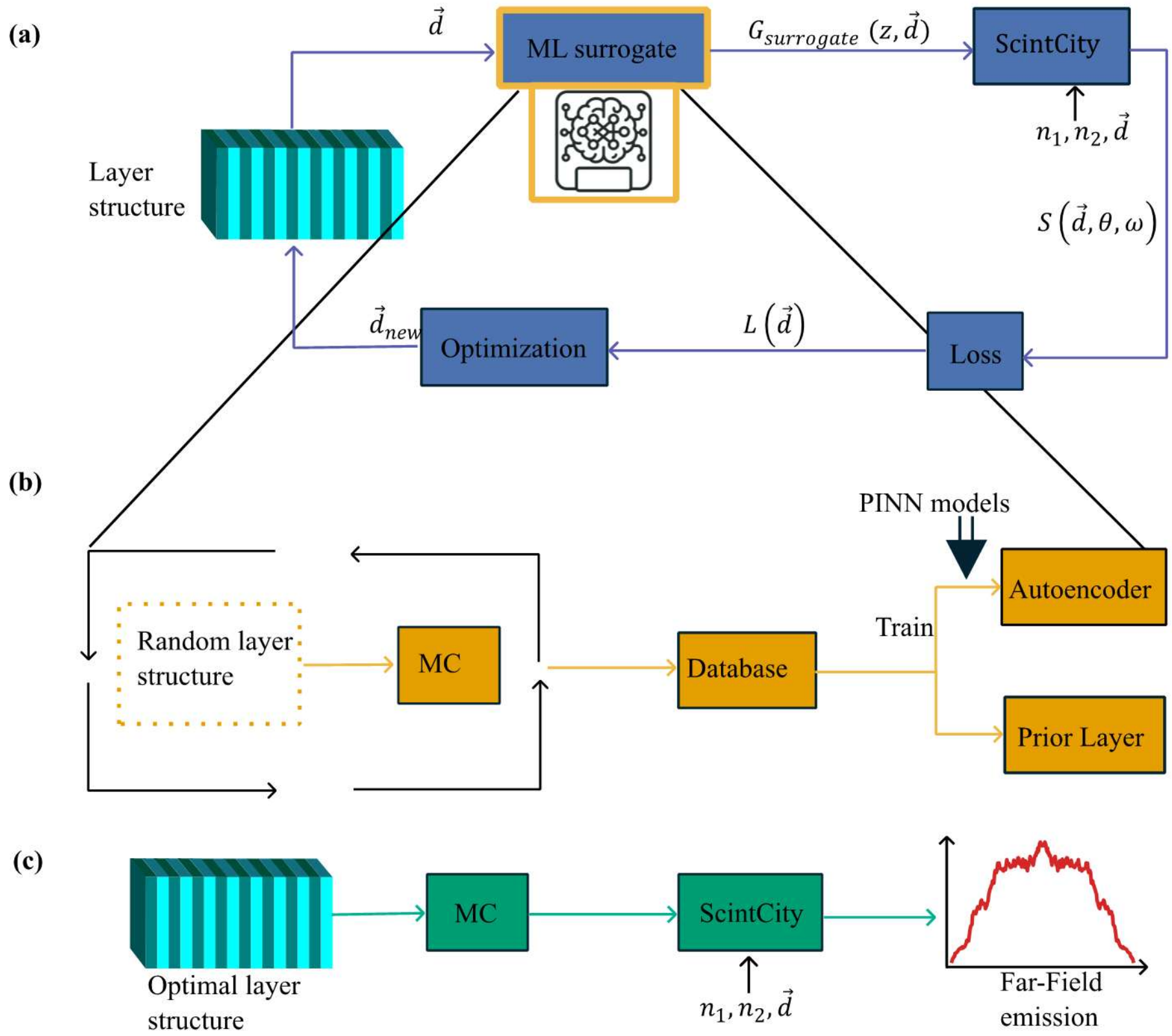

**Figure 2: The scintillator physics-informed neural network (SPINN) design for optimization of layered structures. (a) Optimization process.** The forward model applies the scintillation surrogate module on the layers vector $\vec{d}$, yielding the estimated photon emission density function $G(z, \vec{d})$. Based on this density, ScintCity[27] estimates the emission far-field angular intensity distribution. This intensity is converted into a scalar loss according to the figure-of-merit, used by optimization procedure to update the layers vector. The optimized layer structure is returned upon convergence of the process. **(b) Neural network training.** To train the surrogate model, we begin by running multiple MC simulations for different layered structures at different thicknesses. These results are concatenated into a database on which we train the different surrogate models that we use to predict photon emission density functions. **(c) Evaluation of the scintillator performance.** The performance of every candidate scintillator is tested using the full MC simulation (Geant4[29,40]), followed by the photonic simulator (ScintCity[27]). We then compare the performance of the structure with other candidate scintillators.

The SPINN methodology enables inverse design of nanolayered hybrid scintillator structures. Figure 4(a) outlines the main components of the methodology, implementing the optimization procedure imposed by Eq. (2). The methodology is based on a PINN-inspired learning algorithms that replace the non-differentiable MC simulations. The neural network model learns the spatial distribution of emitters in the layered structure from the thicknesses of the layers, according to the procedure in Figure 4(b). Once trained, the model estimates the spatial distribution of emitters for a given structure and replaces the MC simulation, making the computation of $G\left(z, \vec{d}\right)$ accurate and efficient. This surrogate model approach[39] solves the issues of MC integration: intensive computations, and differentiability, as the neural network surrogate model calculations are faster than the heavy MC simulations and can be accelerated by appropriate hardware. Additionally, gradient calculation of these neural networks is relatively straightforward[10] and easy to implement, as we demonstrate below[41] Following the design, Figure 4(c) shows the evaluation process of the optimized nanolayer structure, where the emission characteristics are evaluated using the MC simulation, independently of the surrogate model.

## The physics-informed neural network (PINN) surrogate architecture

The scintillation surrogate model is designed to replace the cumbersome MC module in the SPINN architecture. We demonstrate this concept on nanolayered hybrid structures, alternating a scintillator and a dielectric material. The model predicts the emitted photons density as a function of the vector of layers' thicknesses, $\vec{d}$, representing the layered structure. Cylindrical symmetry is used to reduce the dimensions of the predicted emitted photon density from 3 dimensions (Cartesian coordinates) to cylindrical coordinates ($r = \sqrt{x^2 + y^2}, z$) averaging out the azimuthal distribution. The output is thus encoded by a 2D histogram of the number of photons emitted at each point in space.

The first surrogate model we consider is architecturally simple: a 3-layer neural network with full connectivity between the layers, trained to directly map the vector of layer thicknesses to the emission distribution. This, however, performs poorly as highly noisy data and large parameters of the model are prone to overfitting and fails to capture the true physical process that occur in scintillation.

To improve performance, we introduced prior knowledge about the emission in the structures by explicitly modeling the emission profile shape, to steer the training process into desired solutions. The prior layer model assumes the layers are thin enough such that the emission within each layer is uniform in the $z$ direction (perpendicular to the structure's layers). We introduce the radial coordinate into the framework (despite the target angular emission optimization depending solely on $z$) to ensure compatibility with future inverse design objectives that may require radial dependence. The radial distribution is assumed to decay exponentially. For layer $i$ in the structure, the emission is modeled as:

$$G_i(z, r) = A_i e^{-r/B_i} \tag{3}$$

where $A_i, B_i$ are the relevant coefficients for layer $i$, and r is the radial coordinate with respect to the axis of the photonic structure. The model then learns the relation between the layer vector and the weights in each layer, which we can then use to find the density of the emission. Given an input layer vector $\vec{d}$ and a binary flag $b$ representing whether the structure starts with a scintillating layer or dielectric layer, the prior layer model output is as follows:

$$\begin{pmatrix} A_i \\ B_i \end{pmatrix}_{i \in (0,1\ldots,N\}} = \mathrm{PINN}(\vec{d}, b) \tag{4}$$

where $N$ represents the number of layers of the structure. The advantage of this modeling lies in the reduction of learnable degrees of freedom. However, this imposes assumptions on the way the emission behaves, hindering the ability to capture novel physical behaviors outside of the assumed model.

To overcome these limitations, we generalize the modelling with a learned autoencoder approach. We consider $K$ parameters in each of the $N$ layers that hold the information on the emission process, without any assumption about how they control it. To extract these parameters, we use an autoencoder model composed of an encoder that encodes the emission distribution into $N \times K$ parameters and a decoder that decodes $N \times K$ parameters into the emission grid. We train the autoencoder to extract these parameters from a general emission simulation with self-supervised learning. The autoencoder works by taking simulated 2D histograms and passing them through the encoder module to retrieve their latent vectors. In return, these vectors are fed to the decoder module that uses them to reconstruct the 2D histogram. The reconstructed histogram is compared to the simulated histogram using MSE metric. The encoder takes the input data and passes it through a fully connected layer and activation functions to get the latent space parameters. Similarly, the decoder passes the latent space parameters through a fully connected layer and activation functions to get the reconstructed data. Further information on models' architecture and training parameters can be found in the supplementary information section S1.

In the subsequent stage of the algorithm, we learn the mapping between the physical layer structure and these discovered latent parameters. Once the autoencoder is trained, we generate a dataset where structural labels are passed through the encoder to retrieve their corresponding latent vectors. We then train a second, separate neural network to predict these $K$ parameters directly from the structural input. The primary advantage of this decoupled approach is that the latent space compacts the complex physics into a minimal set of dimensions ($K \approx 5$), making the optimization significantly more efficient than learning the entire high-dimensional structural response in a single step. While increasing $K$ would cause the model to approach a standard fully connected regime, our constrained parameter set ensures the model remains computationally lean and focused on the dominant physical features. During the inference stage, the process is reversed: the second network predicts the latent parameters for a given structure, which are then fed through the pre-trained decoder to reconstruct the full emission map. This architecture remains agnostic toward the emission physics, avoiding assumptions like exponential decay. The architecture thus provides a more universal, albeit data-intensive, alternative to conventional analytical approximations.

The models were trained on a database consisting of 5,000 MC simulated samples, each one with a random width between 3.9 μm to 4.7 μm. Each simulation consisted of 100,000 8 keV X-Ray photons interactions with the material and were used to create a 2D map of the emitter's distribution. The acquisition of this MC database took around 1 hour to create on 40 cores of Intel Xeon Gold 6230. Further information on simulation settings, geometries and physics can be found in in the supplementary information section S2.

On this database, we trained two models with different training loss functions and compared their performances to the ground truth based on the MC simulation (Fig 3(a, b)).

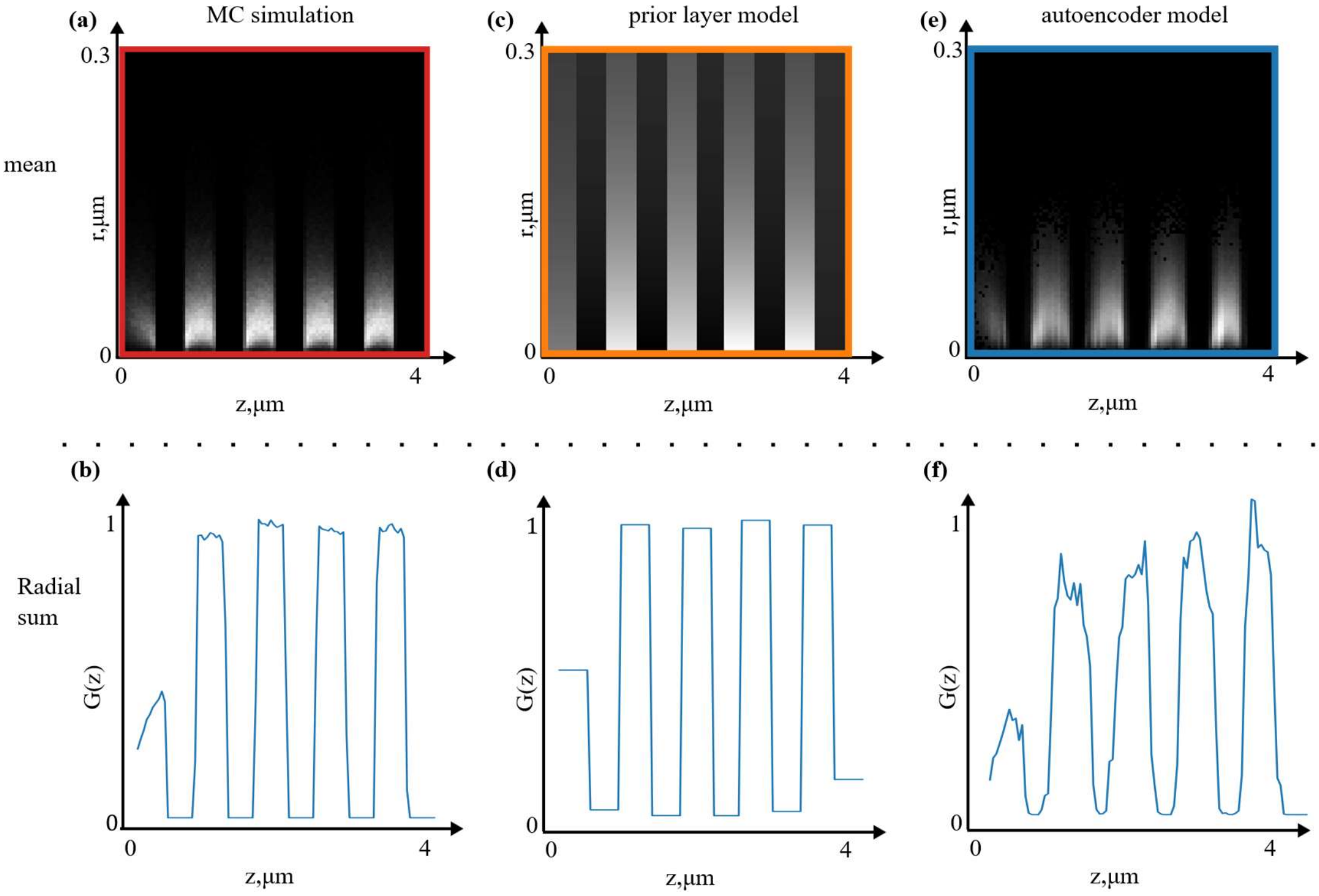

**Figure 3: The scintillation surrogate model of SPINN**, performance evaluation. Top row: Spatial distribution of photon emitters from MC simulations, showing the mean emission density - $G_{\vec{d}}(z,r)$. Bottom row: The sum over the radial coordinates $\bar{G}(z) = \sum_r G_{\vec{d}}(z,r)$. Comparing the MC simulation, which is closest to the ground truth (a,b), with the prior layer model (c,d), and the autoencoder PINN model (e,f). While the autoencoder model is noisier, and fails to capture the smooth transition between layers, it captures well several features that deviate from the prior layer model, including inter-layer changes (like the rise in the first layer), and the curved "onion" dip in emission near the $z$ axis (that appears also in the simulations results).

To demonstrate SPINN, we optimize the emission for X-ray incident perpendicular to the layer structure, which is the typical use case of practical detectors as it maximizes their signal. We consider the signal in the detector as averaged over time with the positions of impacting X-ray photons distributed randomly in the x,y direction (i.e., energy absorption distribution uniform along the radial axis). We calculate the averaged response by summing over the transverse dimension, as presented in Figure 3(b,d,f)

$$G_{\text{PINN}}(z,\vec{d}) = \sum_r r\ \text{PINN}(\vec{d},b)(r,z). \tag{4}$$

In the next section, we test the performance of the surrogate model by quantifying the desired scintillation properties of the entire structure. For that, we solve for the emission properties, relying on a Maxwell equation solver in the nanostructure geometries. The success of SPINN is then determined not only by the comparison to the MC result (Figure 3) but also by comparing the full physical process end-to-end (Figures 4,5).

**Nanophotonic integration and optimization analysis**

We next explore the parameter space of the nanophotonic scintillator structure to identify configurations that maximize the emitted signal. For this purpose, we apply the entire process (Figure 2) to find optimal structures: going beyond the distribution of photon emitters to the eventual light yield signal $L(\vec{d})$. We then evaluate the performance of each optimized structure using a full simulation of MC and ScintCity for photonic emission (Figure 2(c)).

Due to the complex and highly non-convex nature of the design landscape, a systematic examination of the loss surface is essential. This motivates the visualization shown in Figure 4, which maps the loss as a function of key design parameters and highlights regions of high and low performance. There, we mapped the far-field emission, as estimated by each of the prior layer and autoencoder models as well as the previous state-of-the-art exponential emission model. We evaluated the models by comparing their loss map to a ground truth loss map that was achieved by running MC evaluation branch (shown in Figure 2(c)) for each structure.

The autoencoder accurately reconstructed the far-field signal only within a narrow, central region of the parameter space, specifically where the scintillator and dielectric layer thicknesses were nearly equivalent. The sharp decline in accuracy outside this central region suggests a failure in generalization capability. This is attributed to a significant sampling bias in the training dataset, which was overwhelmingly dominated by samples possessing similar layer thickness ratios. The model, therefore, struggled to extrapolate the physical relationships to structurally dissimilar configurations. In sharp contrast, the prior layer model maintained high predictive accuracy across the entire parameter space, consistently yielding the most accurate predictions among the three models. We posit that the incorporation of explicit physical or structural constraints during its training regime acted as an effective regularization mechanism. These constraints prevented the model from overfitting to noise or spurious correlations in the training data, thereby ensuring robust performance on unseen, structurally diverse samples.

The exponential model demonstrated a systematic overestimation of the signal in regions characterized by very low photon yield. This bias arises because the exponential model relies on a macroscopic stopping power parameter to infer light yield per layer. For very thin scintillator layers (on the μm scale), the energy deposition and resulting light yield exhibit non-linear behavior that deviates significantly from the bulk, macroscopic approximation. In regions of high signal, this non-linearity is masked by the dominating effect of structural far-field amplification (the primary goal of the device). However, in regions of weak signal, the uncorrected non-linear effects become the dominant source of error, leading to the observed systematic overestimation.

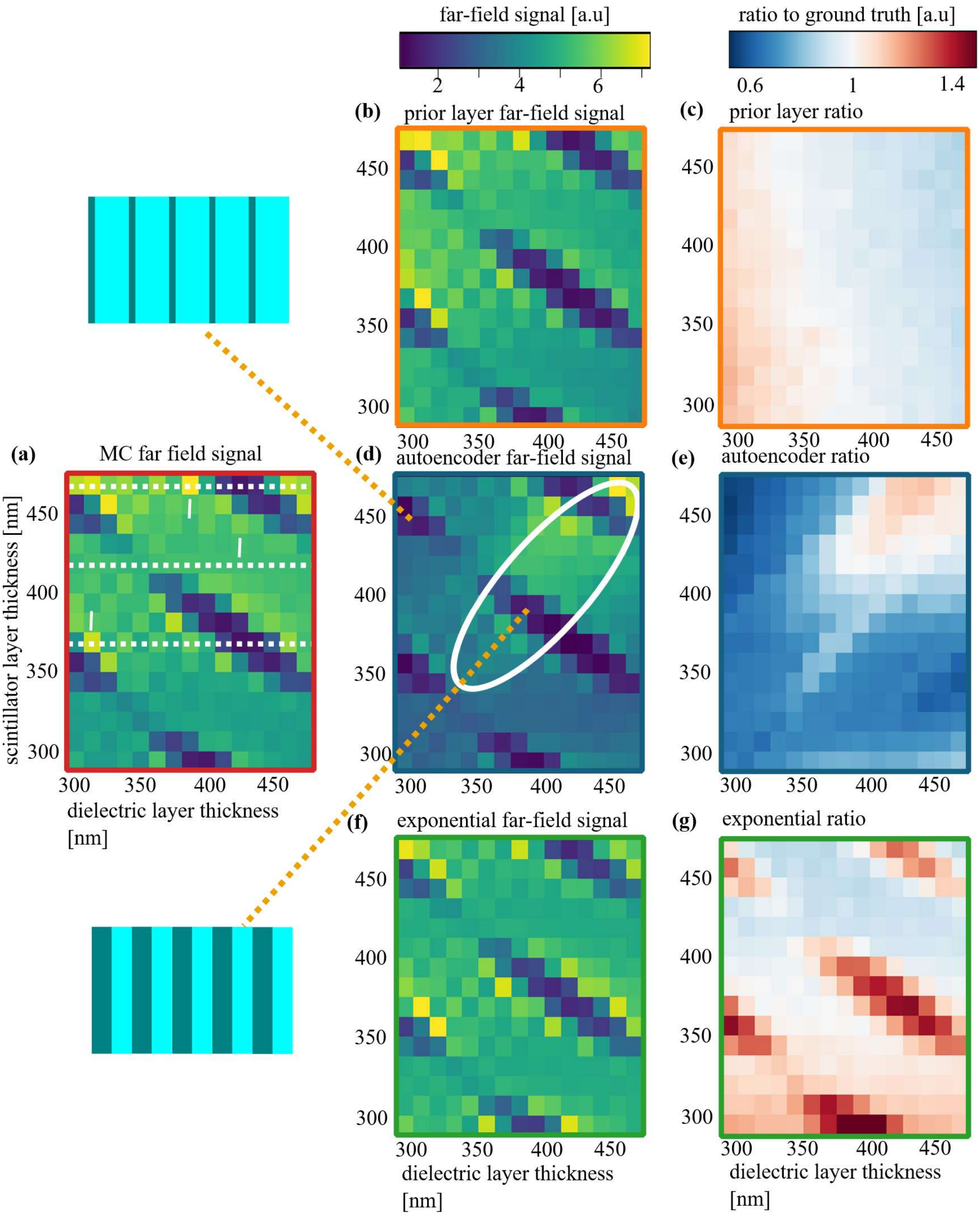

**Figure 4: Analysis of the optimization process,** exemplified on periodic layered scintillators. Scintillator performance (total far-field signal) summarized in a map as a function of the thicknesses of the two layers in the periodic structure. **(a)** Ground truth MC simulation as shown in Figure 2(c). Dotted horizontal lines represent optimization constraints i.e., constant scintillator thickness. white arrows denote the points of highest signal, indicating that the optimal signal is not monotonic with scintillator thickness and in certain contexts can even decrease for thicker structures. **(b,c)** Prior layer forward model performance and ratio to MC ground truth. **(d,e)** Autoencoder model and its ratio to the MC ground truth. Different regions of the parameter space are visualized on the left side. **(f,g)** The previous state-of-the-art exponential model[27] and its ratio to the MC ground truth. The autoencoder model is the best performing in Figure 3, but here under performs, only capturing the far field signal accurately around the $y = x$ axis (scintillator thickness similar to dielectric thickness). Examining the exponential model ratio to the ground truth, we see overestimation of the signal in low intensity regions. This is the indication of the exponential model shortcoming - non proportionality effects cause a major deviation in the number of photons generated.

We optimize the nanophotonic scintillator structure using gradient-based optimization. We evaluated the performance of optimization using different models and using the same initial conditions. The prior layer model demonstrated superior performance, producing results that were closest to the ground truth (simulations) and maintaining its accuracy even when applied outside the original training regions. In contrast, the pure exponential model significantly overshoots the signal emission observed for low yield samples. In these low-yield regions, the signal is not amplified by the layer structure, and the total output is dictated by the sample's bulk yield. The likely cause for the pure exponential model's overshooting is the non-proportionality effect present in the thin samples, which results in an effective yield that is smaller than the macroscopic yield and mean free path parameters used for model input.

To evaluate the quality of the resulting designs, we calculated a ground truth for the scintillator far-field performance directly using the MC simulation coupled to the ScintCity photonic simulator. These results were benchmarked against the current state-of-the-art exponential layer model, commonly used in nanophotonic scintillator optimization as shown in Figure 5(a,b). Despite the autoencoder model performing best in predicting the distribution of emission points (Figure 3), it underperforms relative to the prior layer model because the autoencoder requires a large, high-variance database to remain immune to overfitting, which was not feasible in our setup. Generating such an extensive dataset via MC simulations is computationally intensive. Furthermore, the inherent noise in the MC data is a function of the number of simulated particles, with the statistical uncertainty (or noise) decreasing at a rate of $1/\sqrt{N}$ (where $N$ is the number of simulated particles). In our case, the resulting data contained a level of stochastic noise that the autoencoder struggled to distinguish from actual physical patterns. While the prior layermodel remained stable by relying on its pre-defined structure, the autoencoder's high sensitivity caused it to "learn" this noise, leading to the observed drop in predictive accuracy.

Beyond emission intensity, we also assessed the generalization power of each optimized structure - as demonstrated in Figure 5(a). Moreover, we evaluated their total scintillation output across a range of sample thicknesses, testing how robust the designs remain when applied outside the geometric conditions used during optimization (Figure 5(b)). The comparison of different optimization methods was done using the evaluation branch presented in Figure 2(c). Optimal structures were simulated directly to estimate the total signal.

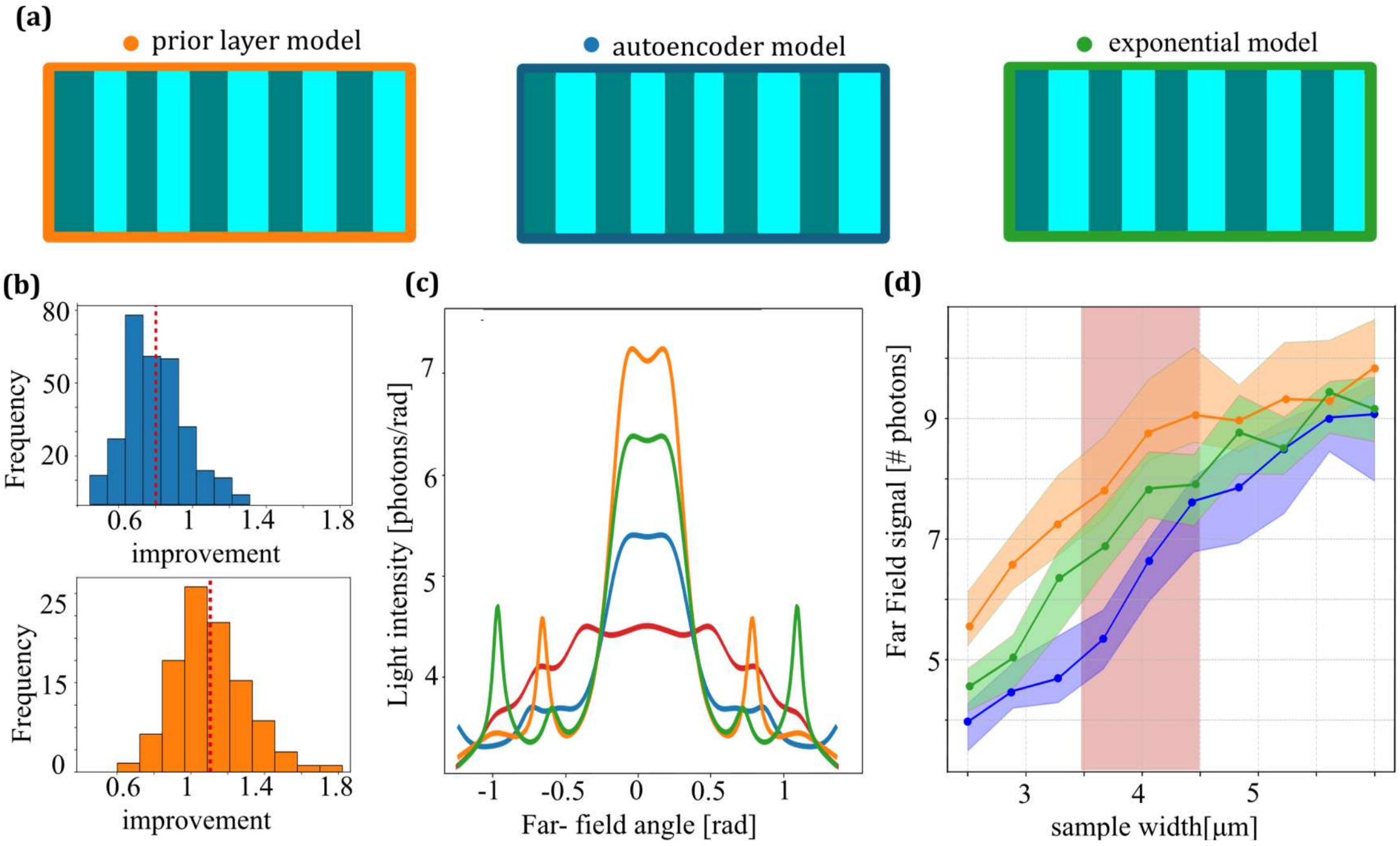

**Figure 5: Example SPINN optimization results**. **(a)** Optimized layered scintillators, found using the different far field models. **(b)** improvement ratio histogram between new models and exponential model. The histogram was calculated by preforming optimization using 3 models on the same initial layer structure and compared the resulting far field signal. The autoencoder (blue histogram) model overall performed worse than previous model with median optimized signal (red dotted line) being 85% of exponential model. The layered structure (orange histogram) improved the median optimized by 15%. **(c)** Comparison of the far field angular intensity distribution for each scintillator design. The intensity distributions are computed using MC and ScintCity (without any surrogate model), to verify the success of the SPINN approach. The optimized structures are compared to a random layered scintillator (red). All comparisons were done using the MC branch presented in Figure 2(c) The previous state-of-the-art method of optimization is the exponent model (green), bypassing a random layered scintillator by a factor of 2 lower than the factors of achieved by our SPINN optimized scintillators. The autoencoder model improved the far field signal by a factor of 2.2 and prior layer produced the best result of 2.8 improvement factor. **(d)** The generalization of the optimization shows that the prior

layer outperformed both other models, surpassing them even outside the training region database (red rectangle). Shaded regions provide a statistical representation of 30 independent optimization runs, where the central line tracks the median value and the enveloped area marks the 25th–75th percentile range to ensure robustness against outliers. We can see that for thick samples, model performance generally converges, while for very thin samples, the improvement of the prior layer signal increases compared to the exponential model. This is likely due to the non-proportionality effects that we discussed previously.

## Discussion and outlook

The SPINN methodology provides a pathway for the inverse design of scintillators, addressing the long-standing incompatibility between traditional MC simulations and gradient-based optimization methods. By replacing computationally intensive and non-differentiable MC simulations with a neural network surrogate model, SPINN enables efficient optimization of layered scintillator structures.

While the autoencoder provided better solutions for the distribution of emission points, its overall utility was severely limited once used for optimizing the overall scintillator far-field signal. In fact, it underperformed relative to the previous-state-of-the-art exponential model, while the prior layer model outperformed it by 15%. The reason is probably the limited dataset relative to the space of parameters, which prevented the model from locating the true global minimum across the full range of input parameters, restricting its efficacy to the densely sampled regions of the training dataset.

The superior performance of the autoencoder model relative to the prior layer model in radial distribution reconstruction suggests that the exponential decay approximation, previously used to describe the radial dependence, fails to accurately capture the underlying physics. While this discrepancy was not critical for the current study—which focused primarily on z-dependence—future applications requiring high-fidelity radial modeling should employ a physical model capable of capturing these complex patterns.

The most significant improvement in emission was achieved for thin structures. For these structures, cross-layer electron interactions become more relevant, and macroscopic linear yield models begin to fail. This was demonstrated in Figure 4 (c), where the exponential model overestimates the signal in low yield regions.

The apparent fluctuations in the signal-versus-thickness curve presented in Figure 5(d) are attributed to thin-film interference effects within the layered architecture. While the global trend indicates an increase in total emission with thickness, the local response is non-monotonic due to phase-dependent phenomena. As shown in Figure 4, the signal does not scale linearly with sample volume; rather, specific thickness increments induce destructive interference, which suppresses the out-coupling of light to the far field. Consequently, a reduction in layer thickness can paradoxically enhance the observed signal by optimizing the extraction efficiency. This oscillatory behavior becomes less pronounced as the number of layers increases—owing to the expanded optical degrees of freedom—yet remains a distinct feature in our 10-layer optimization. It is critical to note that these variations are not numerical artifacts of the simulation or optimization parameters, but are inherent optical characteristics of the scintillator structure.

Future investigations should focus on refining model architectures, for instance, by integrating physics-informed loss functions into the autoencoder. More advanced optimization techniques could improve the performance of SPINN further, especially with the autoencoder architecture (e.g., using adaptive learning rates). Such efforts could improve prediction accuracy, particularly in the poorly sampled regions of the parameter space. Alternatively, the implementation of Bayesian or stochastic neural networks offers a path for integrating simulation noise directly into the model. By capturing statistical properties that more closely mirror the underlying physical processes, this approach enables the optimization of high-dimensional parameter spaces, including those requiring the calculation of higher-order moments of the emission distribution. The SPINN design remains modular and can be generalized to multi-material, 2D/3D scintillators, with alternative performance metrics such as timing resolution or spectral selectivity.

Another emerging frontier that could benefit from the SPINN methodology is quantum-optical scintillator design. Recent advances in this field[19,23] aim to enhance scintillation performance by engineering collective emission modes, transforming spontaneous emission from an incoherent sum of emitters into a cooperative quantum process. These systems are usually characterized by small dimensions and complex geometries that cannot be captured by standard macroscopic optimization strategies but instead fit well to SPINN and its application for nanophotonic scintillators that we demonstrate here. Future work could couple such quantum-optical simulations—accounting for collective decay, mode hybridization, and photonic correlations—with SPINN's differentiable transport and inverse-design architecture. This integration may pave the way toward quantum-enhanced scintillators, where the interplay between (nano)photonic environment and quantum collective behavior outperforms the long-established classical limits of efficiency and resolution.